\documentclass{lf}

\newcommand{\lsim}{\raisebox{-0.5mm}{$\stackrel{<}{\scriptstyle{\sim}}$}}

\newcommand{\qsq}{\mbox{$Q^2$}}

\begin{document}

\title{Deeply Virtual Compton Scattering at H1 and ZEUS}

\author{Laurent Favart}

\address{ {\rm On behalf of the H1 and ZEUS Collaborations} \\
Universit\'e Libre de Bruxelles, Belgium\\
E-mail: lfavart@ulb.ac.be}


\maketitle

\abstracts{
Results on Deeply Virtual Compton Scattering
at HERA measured by the H1 and ZEUS Collaborations are presented.
The cross section, measured for the first time, is reported for
$Q^2 > 2\,{\rm GeV}^2$, $30 < W < 120\,{\rm GeV}$ and
$|t| < 1\,{\rm GeV}^2$.}

\section{Introduction}

The study of exclusive final states in electron-proton diffractive
interactions
is a very powerful tool to investigate the applicability and
the relevance of perturbative Quantum Chromo Dynamics (QCD)
 in this field\cite{Abramowicz}.
At HERA, the wide kinematic range in the photon virtuality,
\qsq, provides particular insight
into the interplay between perturbative and non-perturbative regimes in
QCD.
Moreover, a considerable interest comes from the access diffractive
events give to a new class of parton distribution functions,
the skewed parton distributions (SPD)\cite{Diehl}
that can be interpreted as parton correlation functions in the proton.

\begin{figure}[htbp]
 \begin{center}
  \epsfig{figure=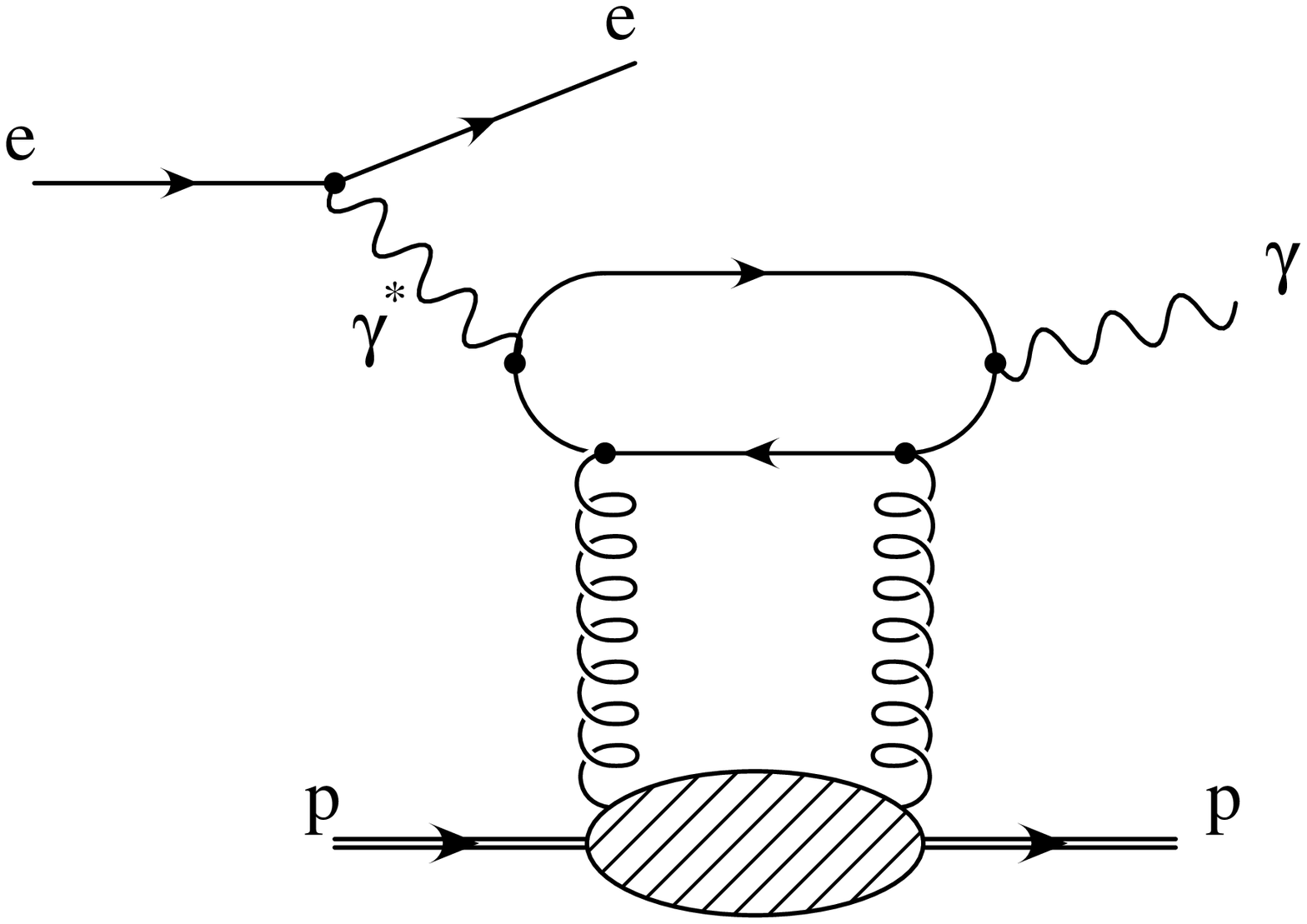,height=0.29\textwidth}
  \hspace*{0.2cm}
  \epsfig{figure=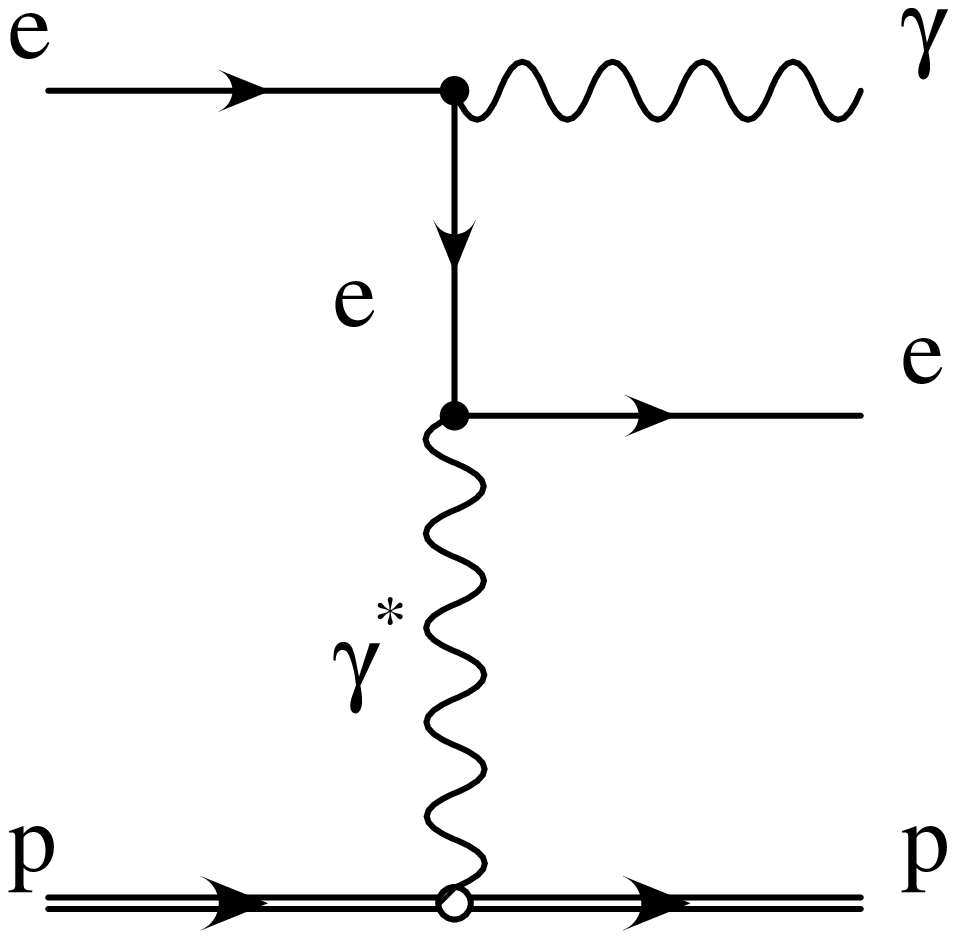,height=0.24\textwidth}
  \hspace*{0.2cm}
  \epsfig{figure=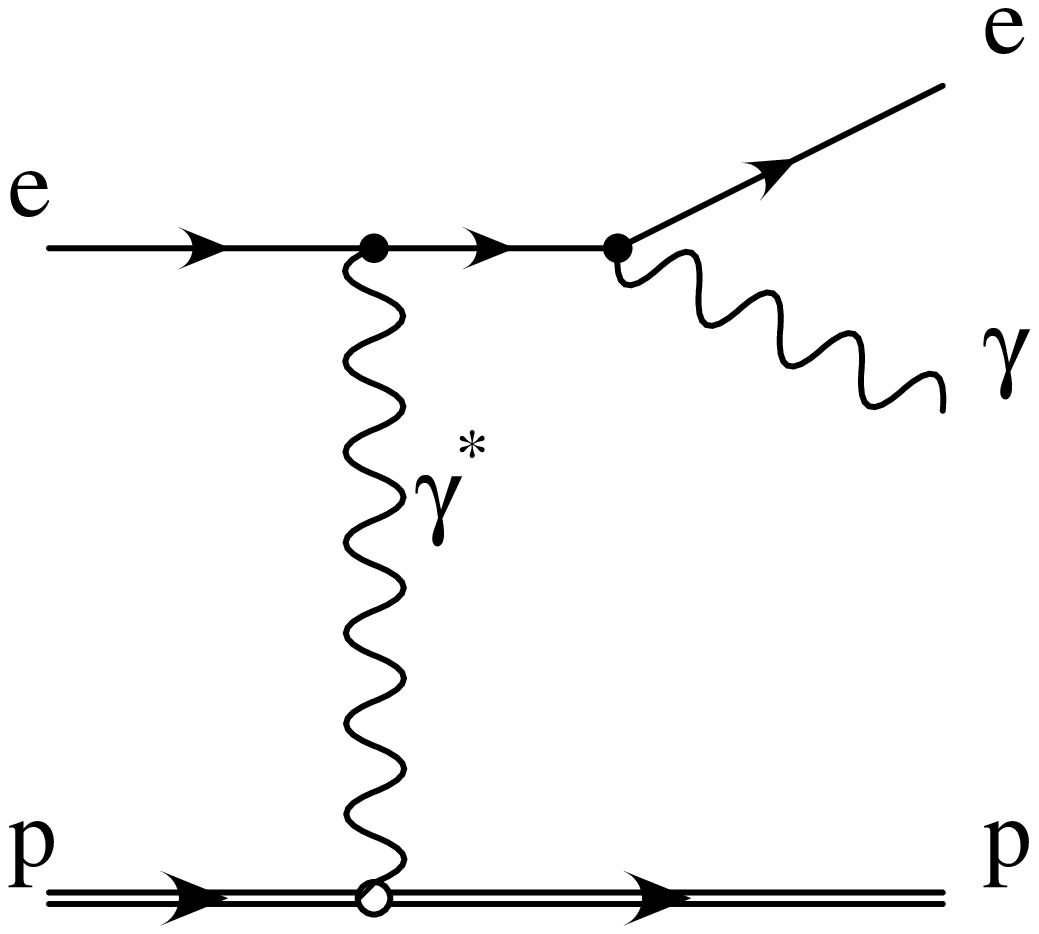,height=0.24\textwidth}
  \\
  \begin{picture}(100,0)
  \put(-44,-1){\bf a)}
  \put(64,-1){\bf b)}
  \put(153,-1){\bf c)}
  \end{picture}
  \caption{The DVCS (a) and the Bethe-Heitler (b and c) processes.}
  \label{fig:diag1}
 \end{center}
 \vspace*{-0.5cm}
\end{figure}

Here we report the first results on the
Deeply Virtual Compton Scattering (DVCS) (Fig.~\ref{fig:diag1}a) i.e.~the
diffractive scattering of a virtual photon off a
proton:
\begin{equation}
 e^{+} + p \rightarrow e^{+} + \gamma + p .
\label{eq:reac}
\end{equation}
This reaction is dominated by the purely electromagnetic
Bethe-Heitler (BH) process (Fig.~\ref{fig:diag1}b and c) whose
cross section,
depending only on QED calculations and proton elastic form factors, is
precisely known and therefore can be subtracted.

Compared to vector meson production, DVCS is theoretically simpler
because the composite meson in the final state is replaced by
the photon, thus avoiding large uncertainties due to the
unknown meson wave functions.

In the presence of a hard scale, the DVCS scattering amplitude
factorises\cite{Radyushkin:1997ki,Collins:1999be,Ji:1998xh}
into a hard scattering part calculable in perturbative QCD
and parton distributions which contain the
non-perturbative effects due to the proton structure.
In practice, even at $Q^2$ values above a few GeV$^2$, the perturbative
regime is strongly influenced by non-perturbative effects which have to
be model based.
In the following, HERA measurement are compared to 
the LO prediction of L.\,Frankfurt, A.\,Freund and M.\,Strikman
(FFS)\cite{ffs}. NLO predictions have been presented for the first time
at this workshop\cite{Freund_nlo}. 

\section{Analysis strategy}

Around the interaction region both experiments, H1 and ZEUS, 
are equipped with 
tracking devices which are surrounded by calorimeters.
Since the proton escapes the main detector through the beam pipe only 
the scattered electron and photon are measured. Therefore the
event selection is based on demanding two electromagnetic clusters,
one in the backward and one in the central or forward part of the 
detector ($\theta \lsim 140^o$ - the backward direction 
($\theta=0$) is defined as the direction of the 
incoming electron). If a track can be reconstructed it has to be
associated 
to one of the clusters and determines the electron candidate.
To enhance the DVCS contribution in comparison 
to the Bethe--Heitler process the phase space has to be restricted
by demanding the photon candidate in the
forward part of the detector.
\\

The H1 analysis selects more specifically the elastic component by using, 
in addition, detectors which are placed close to the 
beam pipe and which are used to identify 
particles originating from proton dissociation processes. 

\section{Results}

\subsection{ZEUS}

The first observation of the DVCS process was reported by the ZEUS
collaboration in 1999\cite{saull}.  In the analysis a photon virtuality
$Q^2 > 6\,{\rm GeV}^2$ is demanded. 
In Fig. 2 the polar
angular distribution of the photon candidates is shown. 
A clear signal
above the expectations for the Bethe--Heitler process is observed.
The LO calculation including the DVCS and the Bethe--Heitler processes 
achives a good description of the experimental data.
A clear DVCS signal is still seen after the photon energy cut is
increased.
Also a shower shape analysis of the calorimetric clusters 
was performed that shows that the signal
originates from photons and not from
$\pi^0$ background.
\\

\begin{minipage}[h]{5cm}
 {\footnotesize Figure 2. Distribution (uncorrected) of the polar angle of the
   photon candidate with an energy above $2\,{\rm GeV}$.
   Data correspond to the full circles. The
   prediction for the Bethe--Heitler process is indicated by the
   open triangles. The prediction of Frankfurt et al. based on
   calculations including the Bethe--Heitler and the DVCS process
   is indicated by the open circles.}
\end{minipage}

\begin{minipage}[h]{6cm}
\setcounter{figure}{2}
 \begin{picture}(100,100)
  \put(160,50){\epsfig{figure=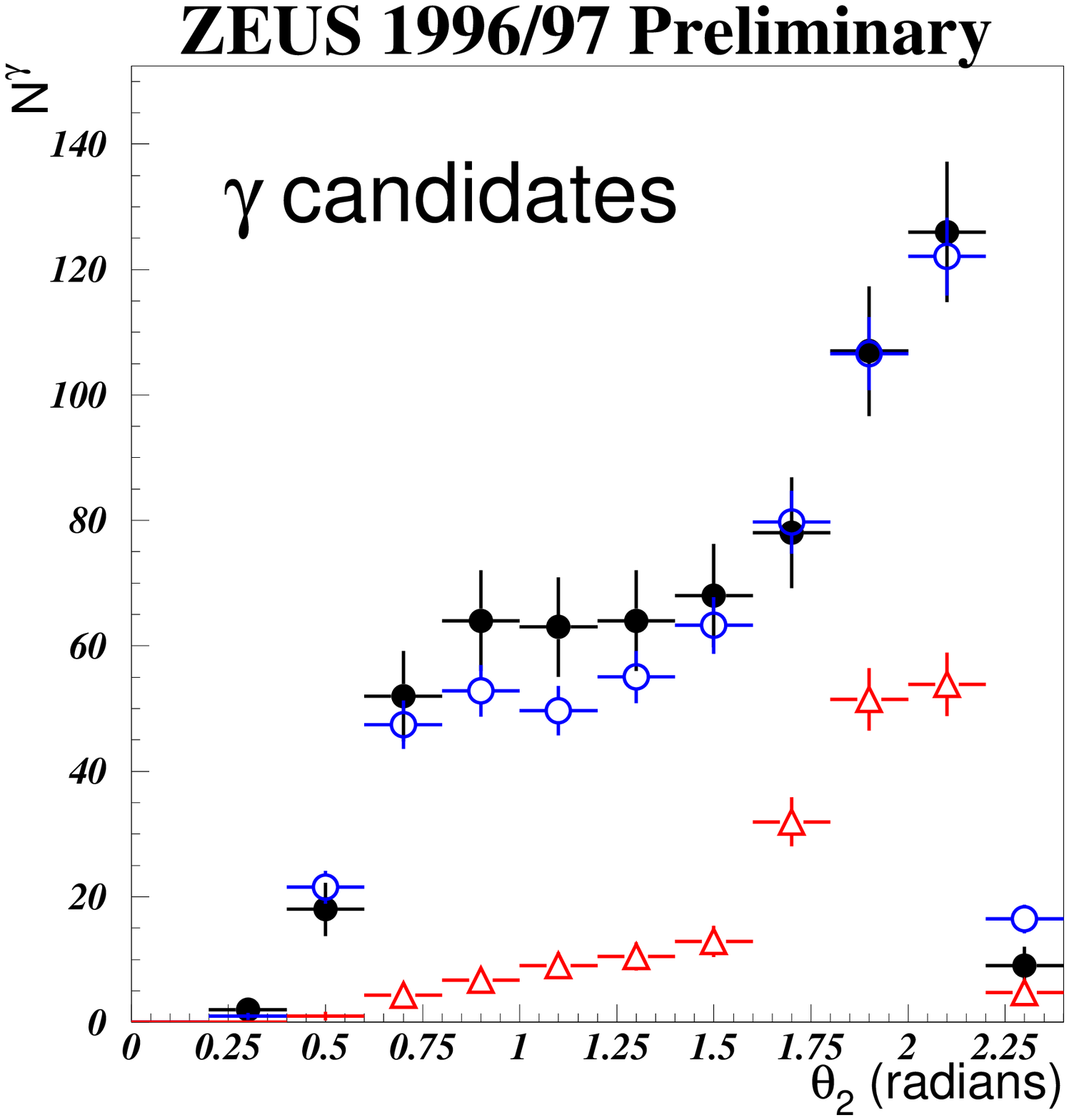,height=5.5cm}}
 \end{picture}
 \vspace{-2.5cm}
\end{minipage}

\subsection{H1}

In the H1 analysis, the DVCS cross section is measured in the
kinematic region:
$2 < Q^2 < 20\,{\rm GeV}^2 $,
$ |t| < 1\,{\rm GeV}^2$
and
$30 < W < 120\,{\rm GeV}$.
The proton
dissociation background has been estimated at around 10\% and subtracted 
statistically assuming the same $W$ and $Q^2$ dependence as for the elastic
component.
The acceptance, initial state radiation of real photons and
detector effects have been
estimated by MC to extract the elastic cross section.
\\

\begin{figure}[htbp]
 \begin{center}
  \epsfig{figure=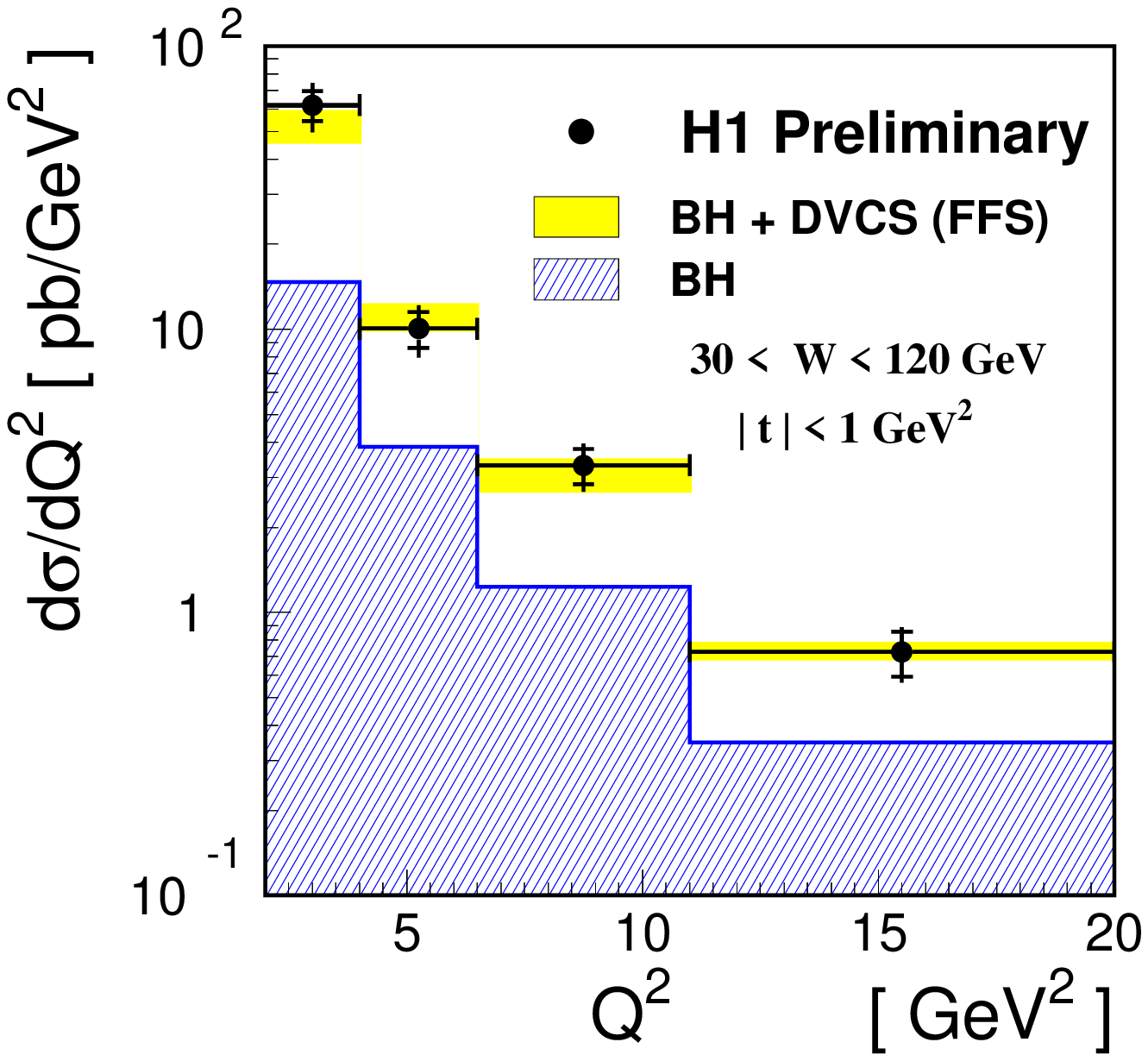,height=5.0cm}
  \epsfig{figure=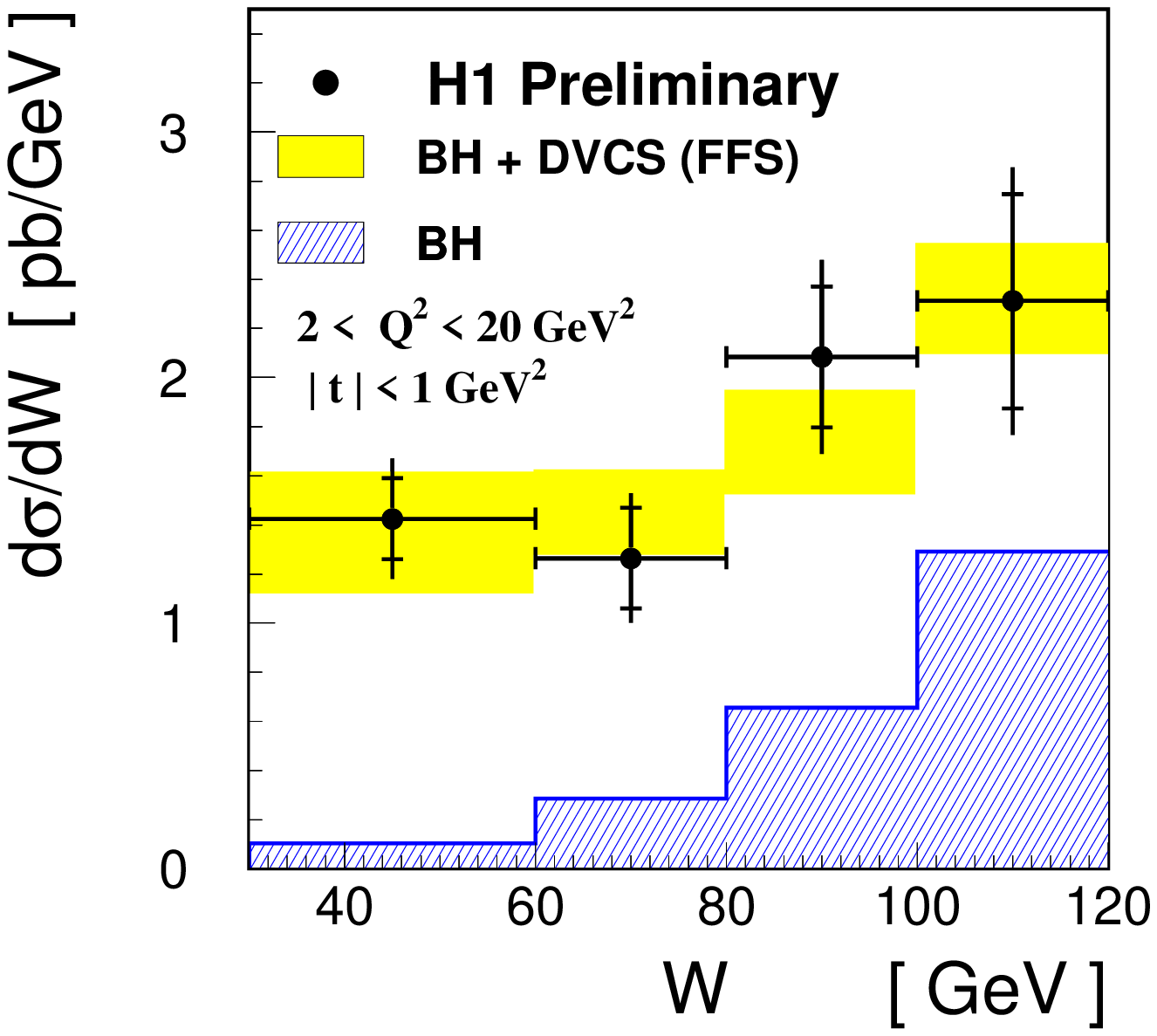,height=5.0cm}
  \caption{The measured cross sections of the reaction
  $e^+ p \rightarrow e^+ \gamma p$ as a function of
  $Q^2$ and $W$ are shown and compared to theoretical predictions.
  The uncertainty in the theoretical prediction, shown here as a shaded
  band is dominated
  by the unknown slope of the $t$ dependence of the DVCS part of
  the cross section, assuming $7 < b < 10\, {\rm GeV}^{-2}$.}
 \label{fig:h1}
 \end{center}
 \vspace*{-0.2cm}
\end{figure}

In Fig.~\ref{fig:h1} the differential cross sections
as a function of $Q^2$ and of $W$ are shown. The data
are compared with the Bethe--Heitler prediction alone and with
the full calculation including Bethe--Heitler and DVCS.
The description of the data by
the calculations is good,
in shape and in absolute normalization when a $t$ slope is chosen between
7 and 10~GeV$^{-2}$ covering the measured
range for light Vector Meson production.
\\

It is important to notice that, at the LO in the leading twist
approximation, the interference term cancels out when
integrating over the azimuthal angle of the final state photon (as in the 
differential cross sections in $Q^2$ and of $W$).

\section{Conclusion}

The DVCS process has been observed by the H1 and ZEUS Collaborations.
The first Cross section measurements have been presented.
The experimental results are well described by the LO
calculations of Frankfurt et al.

\begin{flushright}
{\small
The author is supported by the Fonds National \\
de la Recherche
Scientifique of Belgium.}
\end{flushright}
\end{document}